\documentstyle[aps,preprint]{revtex}

   \font\tenmsb=msbm10 scaled\magstep 1
   \font\sevenmsb=msbm7 scaled \magstep 1
   \font\faivemsb=msbm5 scaled \magstep 1
\newfam\msbfam
      \textfont\msbfam=\tenmsb
      \scriptfont\msbfam=\sevenmsb
      \scriptscriptfont\msbfam=\faivemsb
\def\Bbb#1{{\fam\msbfam #1}}
\font\tengothic=eufm10 scaled\magstep 1
\font\sevengothic=eufm7 scaled\magstep 1
\newfam\gothicfam
      \textfont\gothicfam=\tengothic
      \scriptfont\gothicfam=\sevengothic

\newcommand{\be}{\begin{equation}}
\newcommand{\ee}{\end{equation}}
\newcommand{\dlt}{\delta}
\newcommand{\Dlt}{\Delta}
\newcommand{\ra}{\rightarrow}
\newcommand{\vp}{\varphi}
\newcommand{\bt}{\beta}
\newcommand{\al}{\alpha}
\newcommand{\prt}{\partial}
\newcommand{\dgr}{\dagger}
\newcommand{\Om}{\Omega}
\newcommand{\om}{\omega}
\newcommand{\Lbd}{\Lambda}
\newcommand{\lbd}{\lambda}

\newcommand{\sgm}{\sigma}

\newcommand{\cH}{{\cal H}}
\newcommand{\cY}{{\cal Y}}
\newcommand{\bk}{{\bf k}}
\newcommand{\br}{{\bf r}}
\newcommand{\bp}{{\bf p}}
\newcommand{\ba}{{\bf a}}

\tightenlines

\begin{document}

\draft

\title{Mesoscopic Phase Separation in Anisotropic
Superconductors}
\author{V.I. Yukalov$^{1,2}$ and E.P. Yukalova$^{2,3}$}
\address{
$^1$Bogolubov Laboratory of Theoretical Physics, \\
Joint Institute for Nuclear Research, Dubna 141980, Russia\\ [5mm]
$^2$ Freie Universit\"at Berlin, Fachbereich Physik,
Institut f\"ur Theoretische Physik, \\
WE 2, Arnimallee 14, D-14195 Berlin, Deutschland \\ [5mm]
$^3$Department of Computational Physics, Laboratory of Information
Technologies, \\
Joint Institute for Nuclear Research, Dubna 141980, Russia}

\maketitle

\vskip 1cm

General properties of anisotropic superconductors with mesoscopic phase
separation are analysed. The main conclusions are as follows: Mesoscopic
phase separation can be thermodynamically stable only in the presence of
repulsive Coulomb interactions. Phase separation enables the appearance
of superconductivity in a heterophase sample even if it were impossible
in pure-phase matter. Phase separation is crucial for the occurrence of
superconductivity in bad conductors. Critical temperature for a mixture
of pairing symmetries is higher than the critical temperature related
to any pure gap-wave symmetry of this mixture. In bad conductors, the
critical temperature as a function of the superconductivity fraction
has a bell shape. Phase separation makes the single-particle energy
dispersion softer. For planar structures phase separation suppresses
$d$-wave superconductivity and enhances $s$-wave superconductivity.
These features are in agreement with experiments for cuprates.

\vskip 1cm


\newpage

\section{Introduction}

It is generally accepted that the majority of high-temperature
superconductors, such as cuprates, possess two principal properties
distinguishing them from the conventional low-temperature superconductors.
These properties are mesoscopic phase separation and  anisotropic pairing
symmetry.

Phase separation in superconductors implies that not the whole volume
of a sample is actually superconducting but it is separated into regions
of superconducting and normal phases. The latter may even be insulating.
There exist numerous experiments confirming the occurrence of the phase
separation in high-temperature superconductors, as is summarized in
Refs. [1--4].

The phase separation is termed mesoscopic since the regions of coexisting
phases form a kind of fog of clusters or droplets, whose typical sizes,
corresponding to the coherence length $l_{coh}$, are in between the mean
interparticle distance $a$ and the length of the sample $L$, so that
$$
a\ll l_{coh} \ll L \; .
$$
These regions are intermixed, being randomly distributed in space. In
general, the phase droplets are not static but can be dynamic, randomly
fluctuating in time. In any case, whether they are static or not, their
main features are the mesoscopic size and chaotic space location. Because
of the random spatial distribution of the mesoscopic phase nuclei, they
can be called heterophase fluctuations [5].

The mesoscopic phase separation is, actually, a very general phenomenon
inherent to condensed matter [4--7]. This phenomenon happens in many
systems, being responsible for a variety of unusual effects. For instance,
it plays the key role in colossal magnetoresistant materials [8--10]
and relaxor ferroelectrics [11,12]. Heterophase fluctuations, spatial
or spatio-temporal, can exist in physical systems, without any noticeable
external influence, thus, being {\it self-organized} [5,13]. The action
of external forces can, of course, provoke the appearance of such
mesoscopic fluctuations [5,14], making them more intensive. However,
in general, the noticeable external perturbations are not compulsory,
and heterophase fluctuations can really  arise in a self-organized way.
In some cases, these fluctuations can be triggered by infinitesimally
small stochastic noise that always exists in all realistic systems,
which are never completely isolated from their surrounding but are not
more than quasi-isolated [5,15].

The possibility of mesoscopic phase separation in superconductors
was advanced [16,17] yet before high-temperature superconductors were
discovered [18]. Theoretical models confirm that this phenomenon can
be thermodynamically profitable, rending the heterophase material more
stable [17,19--23].

Another specific feature of high-temperature cuprate superconductors
is the anisotropy of the gap. A number of experiments point at the
predominantly $d$-wave symmetry of the superconducting order parameter
[24,25], though in some cases one claims that the isotropic $s$-wave
symmetry can be dominant [26--29]. The majority of experiments evidence
the existence of the mixed $s+d$ pairing in cuprates [25,30--38]. Several
theoretical models, blending $s$-wave and $d$-wave features, describe the
$s+d$ superconducting gap state and provide a reasonable explanation for
various experiments [39--52]. Thus, the occurrence of anisotropy in the gap
of high-temperature cuprate superconductors seems to be well established.

In the present paper, we suggest a model of superconductor, which combines
two main features: {\it mesoscopic phase separation} and {\it anisotropic
pairing symmetry}. We study interplay between these characteristics. We
do not narrow down the consideration by fitting parameters to a particular
material, but we rather concentrate on the general properties of the model.
The basic goal of the paper is to formulate the principal qualitative
conclusions characterizing such a superconductor with both phase separation
and an anisotropic gap.

The treatment of the suggested model of superconductor is based on the
theory of heterophase materials [5] possessing the properties typical of
the matter with mesoscopic phase separation. There are some important
points that are worth emphasizing in order to better understand the following
consideration.

First of all, one should keep in mind that the basic spatial structure of
matter is defined by ions forming a crystalline lattice. Charge carriers,
such as electrons and holes, exist inside the given crystalline structure of
a particular solid. Therefore the properties of all spatial characteristics,
e.g. interaction potentials, are prescribed by a concrete crystalline
structure of ions forming the lattice.

Superconductivity or normal conductivity are the features of the charge
carriers, reflecting the level of correlations between the latter. A solid
with the same crystalline structure, as is well known, can be superconducting
or not depending on the values of thermodynamic parameters. The occurrence of
superconductivity of carriers does not substantially change the crystalline
structure of ions. Thus, superconducting and normal phases of carriers may
coexist inside the same crystalline lattice.

The coexistence of different phases, typical of mesoscopic phase separation,
means that the spatial regions of the sample are occupied by different phases,
these regions, generally, have diverse shapes and random spatial locations.
Then, in order to describe the properties of the sample as a whole, one has
to average over phase configurations. The procedure of such a heterophase
averaging is rather nontrivial, being analogous to the renormalization-group
technique, when one averages out one type of fluctuations with temporal or
spatial scales that are distinct from another type of fluctuations. In our
case, the heterophase fluctuations are mesoscopic, which distinguishes them
from microscopic quantum fluctuations. All mathematical details of the 
heterophase averaging over configurations have been thoroughly expounded 
in review [5]. In order that the reader could catch the main points of 
this procedure, these are sketched in the Appendix.

After averaging over phase configurations, one obtains a renormalized
Hamiltonian representing the phase replicas that would occupy the whole sample
with a certain probability. In this way one comes to the picture where all
characteristics do not involve anymore random spatial distributions but
correspond to the averaged quantities resulting from their averaging over
these random phase configurations. In the following sections, we deal with
such averaged characteristics appearing {\it after} the heterophase averaging,
whose essence is surveyed in the Appendix.

\section{Heterophase Superconductor}

A superconductor with phase separation is a sample consisting of intermixed
regions of different thermodynamic phases. Assume there are two phases,
superconducting and normal, enumerated by the index $\nu=1,2$. Let $\nu=1$
correspond to the superconducting phase, while $\nu=2$, to the normal phase.
Quantum states of the phases pertain to the related spaces $\cH_\nu$, which
are the weighted Hilbert spaces [5]. The phases can be distinguished by
their {\it order parameters}, as the gaps $\Dlt_\nu(\bk)$ in the momentum
space, so that
\be
\label{1}
\Dlt_1(\bk) \not\equiv 0\; , \qquad \Dlt_2(\bk) \equiv 0 \; .
\ee
Another way of distinguishing phases is by the associated order indices,
which are defined for reduced density matrices [53] and have been generalized
for arbitrary operators [54]. The {\it order index} for a bounded operator
$\hat A$ is
$$
\om(\hat A) \equiv \frac{\log||\hat A||}{\log|{\rm Tr}\; A|} \; .
$$
Considering, in the place of $\hat A$, $p$-particle density matrices
$\hat\rho_{p\nu}$ of the phases $\nu=1,2$, we have the following [53,54].
For the superconducting phase, the order indices of odd density matrices
are
$$
\om(\hat\rho_{p1}) = \frac{p-1}{2p} \qquad (p=1,3,5,\ldots ) \; ,
$$
and those of even density matrices are
$$
\om(\hat\rho_{p1}) =\frac{1}{2} \qquad (p=2,4,6,\ldots ) \; .
$$
But for the normal phase, the order indices of all reduced density matrices
are zero,
$$
\om(\hat\rho_{p2}) = 0 \qquad (p=1,2,3,\ldots) \; .
$$

Coexisting phases occupy different spatial regions of the sample. These
regions are composed of mesoscopic subregions that are randomly intermixed
in space, forming complicated configurations. For each given configuration,
we can define a locally-equilibrium Gibbs ensemble. Then, since the spatial
phase distribution is random, it is necessary to average over these phase
configurations. This procedure makes the basis of the theory of statistical
systems with mesoscopic phase separation [5]. After averaging over random
phase configurations, we come (see Appendix) to the renormalized Hamiltonian
\be
\label{2}
\tilde H = H_1\oplus H_2
\ee
defined on the fiber space
\be
\label{3}
\cY =\cH_1 \otimes \cH_2 \; ,
\ee
being the tensor product of the weighted Hilbert spaces. The phase-replica
Hamiltonians $H_\nu$ can be written in the form
\be
\label{4}
H_\nu = w_\nu H_\nu^{kin} + w_\nu^2 H_\nu^{int} \; ,
\ee
where $H_\nu^{kin}$ is an operator of kinetic energy, $H_\nu^{int}$ is
an operator describing pair interactions, and $w_\nu$ are phase probabilities
satisfying the conditions
\be
\label{5}
w_1 + w_2 = 1 \; , \qquad 0\leq w_\nu \leq 1 \; .
\ee
The phase probabilities are defined as the minimizers of the thermodynamic
potential
\be
\label{6}
\Om = - T\ln{\rm Tr}_\cY \; e^{-\bt\tilde H} \qquad (\bt T \equiv 1) \; .
\ee
Here and in what follows, $T$ is temperature, $k_B=1$. Setting the notation
\be
\label{7}
w_1 \equiv w \; , \qquad w_2 = 1 - w \; ,
\ee
the minimization condition reads
\be
\label{8}
\frac{\prt\Om}{\prt w} = 0 \; , \qquad
\frac{\prt^2\Om}{\prt w^2} > 0 \; .
\ee
This condition shows when mesoscopic phase separation is profitable, as
compared to a pure system.

The average of an operator $\hat A$ is given as
\be
\label{9}
<\hat A>\; \equiv {\rm Tr}_\cY\; \hat\rho\hat A \; , \qquad \hat\rho
\equiv \frac{e^{-\bt\tilde H}}{{\rm Tr}_\cY\; e^{-\bt\tilde H}} \; ,
\ee
with $\hat\rho$ being the statistical operator. Then the first equation from
condition (8) takes the form
\be
\label{10}
<\frac{\prt\tilde H}{\prt w}> \; = 0 \; ,
\ee
and the second equation yields the inequality, being the {\it condition of
heterophase stability}
\be
\label{11}
\left [ < \frac{\prt^2\tilde H}{\prt w^2}> \; - \bt<\left (
\frac{\prt\tilde H}{\prt w}\right )^2 > \right ] >  0 \; .
\ee
To illustrate the meaning of Eqs. (10) and (11), let us take the Hamiltonian
(2) with the terms (4) and suppose that the kinetic part $H_\nu^{kin}$ and
the interaction part $H_\nu^{int}$ do not depend on $w_\nu$. Let us define
the mean kinetic energy $K_\nu$ and the mean interaction potential $\Phi_\nu$
by the notation
\be
\label{12}
K_\nu \equiv \; < H_\nu^{kin}> \; , \qquad
\Phi_\nu \equiv 2<H_\nu^{int}> \; .
\ee
Then Eq. (10) gives the probability of the superconducting phase
\be
\label{13}
w = \frac{\Phi_2+K_2-K_1}{\Phi_1+\Phi_2} \; .
\ee
From here, since $0\leq w\leq 1$, we get
\be
\label{14}
-\Phi_1 \leq K_1 - K_2 \leq \Phi_2 \; .
\ee
The stability condition (11) yields
$$
\Phi_1 + \Phi_2 > \; \frac{\bt}{N} < \left (
\frac{\prt\tilde H}{\prt w}\right )^2 > \; ,
$$
from where it follows that the necessary condition for the stability of a
phase-separated sample is
\be
\label{15}
\Phi_1 + \Phi_2 > 0\; .
\ee
In that case, phase separation becomes thermodynamically profitable, for which,
as is seen from Eq. (15), the existence of repulsive interactions is compulsory.

\section{Structure of Hamiltonian}

Employing the field representation, we deal with the field operators
$\psi_{s\nu}(\br)$, in which $s=\uparrow,\downarrow$ denotes spin and $\br$
is a spatial vector. Fermi commutation relations are assumed. The kinetic
part has the standard form
\be
\label{16}
H_\nu^{kin} = \sum_s \int \psi_{s\nu}^\dgr(\br) \left [ \hat K_\nu(\br) -
\mu\right ]\; \psi_{s\nu}(\br)\; d\br \; ,
\ee
where $\hat K_\nu(\br)$ is a kinetic transport operator and $\mu$, chemical
potential. The interaction part $H_\nu^{int}$, in general, consists of direct
interactions and of effective interactions due to a kind of boson exchange.
This, for instance, can be the phonon exchange, if one considers the usual
picture based on the Fr\"olich Hamiltonian [55--58]. In principle, one may
consider the exchange by other types of bosons, say excitons, but for
concreteness, we shall keep in mind the conventional phonon picture. For
simplicity, and at the same time for generality, we take for the interaction
Hamiltonian the expression
\be
\label{17}
H_\nu^{int} = \frac{1}{2} \sum_{ss'} \int \psi_{s\nu}^\dgr(\br)
\psi_{s'\nu}^\dgr(\br') \hat V_\nu(\br,\br')\psi_{s'\nu}(\br')
\psi_{s\nu}(\br)\; d\br d\br' \; ,
\ee
where the vertex operator $\hat V_\nu$ models all effective interactions,
direct as well as those caused by boson exchange. The vertex operator is
supposed to be symmetric,
\be
\label{18}
\hat V_\nu(\br,\br') = \hat V_\nu(\br',\br) \; .
\ee

In agreement with condition (1), the anomalous averages for the superconducting
phase are not trivial
\be
\label{19}
<\psi_{s1}(\br)\psi_{s'1}(\br') > \; \not\equiv 0 \; ,
\ee
at least for some spins, while such averages for the normal phase are
identically zero,
\be
\label{20}
<\psi_{s2}(\br)\psi_{s'2}(\br')>\; \equiv 0 \; .
\ee

For crystalline matter with a periodic structure, the field operator can be
expanded over Bloch functions, which, for the single-zone case, writes
\be
\label{21}
\psi_{s\nu}(\br) = \sum_k c_{s\nu}(\bk) \vp_k(\br) \; ,
\ee
with $\bk$ being wave vector. Let us introduce the matrix elements over the
Bloch functions $\vp_k(\br)$, resulting in the transport matrix
\be
\label{22}
t_\nu(\bk,\bp) \equiv (\vp_k,\hat K_\nu\vp_p)
\ee
and the vertex
\be
\label{23}
V_\nu(\bk,\bk',\bp',\bp) \equiv (\vp_k\vp_{k'},\hat V_\nu\vp_{p'}\vp_p)\; .
\ee
The latter, due to Eq. (18), has the symmetry property
\be
\label{24}
V_\nu(\bk,\bk',\bp',\bp) = V_\nu(\bk',\bk,\bp,\bp') \; .
\ee

Invoking expansion (21), the kinetic part (16) transforms to
\be
\label{25}
H_\nu^{kin} = \sum_s \sum_{kp} \left [ t_\nu(\bk,\bp) - \mu\dlt_{kp}
\right ]\; c_{s\nu}^\dgr(\bk) c_{s\nu}(\bp)
\ee
and the interaction term(17) becomes
\be
\label{26}
H_\nu^{int} = \frac{1}{2}\; \sum_{ss'}\; \sum_{kk'} \; \sum_{pp'}
V_\nu(\bk,\bk',\bp',\bp) c_{s\nu}^\dgr(\bk) c_{s'\nu}^\dgr(\bk')
c_{s'\nu}(\bp') c_{s\nu}(\bp) \; .
\ee

To make the problem treatable, let us resort to the Hartree-Fock-Bogolubov
approximation, according to which the four-operator products are expressed as
$$
c_1c_2c_3c_4 = c_1c_2<c_3c_4>+<c_1c_2>c_3c_4 - <c_1c_2><c_3c_4> +
$$
$$
+ c_1c_4<c_2c_3> + <c_1c_4>c_2c_3 -
$$
\be
\label{27}
 - <c_1c_4><c_2c_3> - c_1c_3<c_2c_4> - <c_1c_3>c_2c_4 +
<c_1c_3><c_2c_4> \; ,
\ee
where $c_i$ represents any of the operators $c_{s\nu}(\bk)$ or
$c_{s\nu}^\dgr(\bk)$ and the Fermi commutation relations are assumed.
Also, we shall consider the restricted spaces of quantum states, for
which the Bardeen-Cooper-Schrieffer restriction is valid:
$$
c_{s\nu}^\dgr(\bk) c_{s'\nu}(\bk')= \dlt_{ss'}\; \dlt_{kk'}
c_{s\nu}^\dgr(\bk) c_{s\nu}(\bk) \; ,
$$
\be
\label{28}
c_{s\nu}^\dgr(\bk) c_{s'\nu}^\dgr(\bk')= \dlt_{-ss'}\; \dlt_{-kk'}
c_{s\nu}^\dgr(\bk) c_{-s\nu}^\dgr(-\bk) \; .
\ee
This means that the restricted spaces consist of the wave functions for which
spin and momentum are conserved.

The normal average
\be
\label{29}
n_\nu(\bk) \equiv \sum_s <c_{s\nu}^\dgr(\bk) c_{s\nu}(\bk)>
\ee
is the momentum distribution of particles. Introducing the anomalous average
\be
\label{30}
\sgm_\nu(\bk) \equiv \; <c_{-s\nu}(-\bk) c_{s\nu}(\bk) > \; ,
\ee
conditions (19) and (20) can be rewritten as
\be
\label{31}
\sgm_1(\bk)\not\equiv 0 \; , \qquad \sgm_2(\bk) \equiv 0 \; .
\ee

With the approximations (27) and (28), the Hamiltonian (4) can be
diagonalized by means of the Bogolubov canonical transformation
\be
\label{32}
c_{s\nu}(\bk) = u_\nu(\bk) a_{s\nu}(\bk) +
v_\nu(\bk) a_{-s\nu}^\dgr(\bk) \; ,
\ee
in which
$$
|u_\nu(\bk)|^2 = \frac{1}{2}\left [ 1 +
\frac{\om_\nu(\bk)}{E_\nu(\bk)} \right ] \; , \qquad
|v_\nu(\bk)|^2 = \frac{1}{2}\left [ 1 - \;
\frac{\om_\nu(\bk)}{E_\nu(\bk)} \right ] \; .
$$
Here, the single-particle dispersion is
\be
\label{33}
\om_\nu(\bk) = t_\nu(\bk,\bk) + w_\nu M_\nu(\bk) - \mu \; ,
\ee
with the mass operator
\be
\label{34}
M_\nu(\bk) \equiv \sum_p \left [ V_\nu(\bk,\bp,\bp,\bk) - \;
\frac{1}{2} V_\nu(\bk,\bp,\bk,\bp)\right ] \; n_\nu(\bp) \; ,
\ee
and the excitation spectrum
\be
\label{35}
E_\nu(\bk) = \sqrt{\Dlt_\nu^2(\bk)+\om_\nu^2(\bk)}
\ee
contains the gap
\be
\label{36}
\Dlt_\nu(\bk) =  w_\nu \; \sum_p J_\nu (\bk,\bp)\; \sgm_\nu(\bp) \; ,
\ee
where the effective interaction
\be
\label{37}
J_\nu(\bk,\bp) \equiv - V_\nu(\bk,-\bk,-\bp,\bp) \; .
\ee
Then the Hamiltonian (4) reduces to
\be
\label{38}
H_\nu = w_\nu \; \sum_s\; \sum_k E_\nu(\bk) a_{s\nu}^\dgr(\bk)
a_{s\nu}(\bk) + w_\nu C_\nu \; ,
\ee
with the nonoperator term
$$
C_\nu = \sum_k \left [ \om_\nu(\bk) - E_\nu(\bk) +\Dlt_\nu(\bk)
\sgm_\nu(\bk) - \; \frac{1}{2}\; w_\nu M_\nu(\bk) n_\nu(\bk) \right ] \; .
$$
For the averages (29) and (30), one gets
\be
\label{39}
n_\nu(\bk) = 1 - \; \frac{\om_\nu(\bk)}{E_\nu(\bk)}\; {\rm tanh}\;
\frac{w_\nu E_\nu(\bk)}{2T}\; , \qquad
\sgm_\nu(\bk) = \frac{\Dlt_\nu(\bk)}{2E_\nu(\bk)}\; {\rm tanh}\;
\frac{w_\nu E_\nu(\bk)}{2T} \; .
\ee
These expressions have sense for both the superconducting and normal phases;
however, for the normal phase, according to condition (31), one has
\be
\label{40}
\sgm_2(\bk) = \Dlt_2(\bk) = 0 \; ,
\ee
thence
$$
n_2(k) = \frac{2}{\exp\{\bt w_2\om_2(\bk)\}+1} \; .
$$

Note that the phase probabilities $w_\nu$ enter all equations in a
rather nontrivial way, which will essentially influence the properties of
superconductors with mesoscopic phase separation.

\section{Phase Separation}

In what follows, we shall use the notation (7), writing $w=w_1$, and,
similarly, we shall omit, for the sake of simplicity, the index $\nu=1$ at
all related quantities. For instance, we shall write $\Dlt(\bk)$, $\sgm(\bk)$,
$E(\bk)$, and so on instead of $\Dlt_1(\bk)$, $\sgm_1(\bk)$, and $E_1(\bk)$.

The gap equation (36) can be presented as
\be
\label{41}
\Dlt(\bk) = \frac{w}{2}\; \sum_p J(\bk,\bp) \; \frac{\Dlt(\bp)}{E(\bp)}\;
{\rm tanh}\; \frac{wE(\bp)}{2T} \; .
\ee
Looking for a positive solution for $\Dlt(\bk)$, we see that this is possible
when the right-hand side of Eq. (41) is also positive, which requires that
the effective interaction
\be
\label{42}
J(\bk,\bp) >  0
\ee
be positive in the region of momenta making the main contribution in the
summation of Eq. (41).

Another necessary condition is the condition (15) for the profitability
of phase separation. For the mean interaction potential, defined in Eq. (12),
we find
\be
\label{43}
\Phi_\nu = \sum_k M_\nu(\bk) n_\nu(\bk) - 2 \sum_{kp} J_\nu(\bk,\bp)
\sgm_\nu(\bk) \sgm_\nu(\bp) \; .
\ee
Then the stability condition (15) yields
\be
\label{44}
\sum_\nu\; \sum_k M_\nu(\bk) n_\nu(\bk) > 2 \sum_{kp} J(\bk,\bp)
\sgm(\bk)\sgm(\bp) \; .
\ee
From here, we get the necessary condition
\be
\label{45}
\sum_\nu\; \sum_k M_\nu(\bk) n_\nu(\bk) >  0 \; ,
\ee
which tells that some sufficiently strong repulsive interactions are to
be present in the system. Such natural interactions are, of course, Coulomb
interactions. Thus, we come to the first conclusion:

\vskip 2mm

{\it Mesoscopic phase separation in superconductors can be thermodynamically
stable only in the presence of repulsive Coulomb interactions}.

\vskip 2mm

Now let us recall that, as is discussed in the Introduction, the gap of
the hole-doped high-temperature cuprate superconductors displays strong
anisotropic dependence on momentum. To describe the anisotropy, one may
introduce a basis $\{\chi_i(\bk)\}$ of functions $\chi_i(\bk)$ characterizing
the lattice symmetry, with the index $i=1,2,\ldots$ enumerating irreducible
representations of the symmetry group. Let such a basis be defined, being
orthonormal and complete,
$$
\sum_k \chi_i^*(\bk)\chi_j(\bk) =\dlt_{ij} \; , \qquad
\sum_i \chi_i^*(\bk) \chi_i(\bp) = \dlt_{kp} \; .
$$
Then one may expand (see [25], [59]) over this basis the effective interaction
\be
\label{46}
J(\bk,\bp) = \sum_{ij} J_{ij} \chi_i(\bk)\chi_j^*(\bp)
\ee
and the gap
\be
\label{47}
\Dlt(\bk) = \sum_i \Dlt_i \chi_i(\bk) \; .
\ee
Using this, the gap equation (41) reduces to
\be
\label{48}
\Dlt_i = \sum_j A_{ij} \Dlt_j \; ,
\ee
where
\be
\label{49}
A_{ij} \equiv \sum_p \frac{wJ_{ij}}{2E(\bp)} \; {\rm tanh}\left [
\frac{wE(\bp)}{2T}\right ] \; \chi_i^*(\bp) \chi_j(\bp) \; .
\ee
The system of uniform algebraic equations (48) possesses nontrivial solutions
when
\be
\label{50}
{\rm det}(\hat 1 - \hat A) = 0 \; ,
\ee
where $\hat 1=[\dlt_{ij}]$ is the unity matrix and the matrix
$\hat A=[A_{ij}]$ is composed of elements (49).

The effective interaction (46) consists of an attractive part, caused by
phonon exchange, and  a repulsive part, due to direct Coulomb interactions
[60], because of which $J_{ij}$ has the structure
\be
\label{51}
J_{ij} = \left ( \frac{|\al|^2}{\tilde\om_0^2}\; - M_0
\right ) \; b_{ij} \; ,
\ee
in which $\al$ is the charge-lattice coupling, $\tilde\om_0$ is the
characteristic lattice frequency in the presence of heterostructural
fluctuations [61,62], connected by the relation
\be
\label{52}
\tilde\om_0 =\sqrt{w}\; \om_0
\ee
with the characteristic lattice frequency $\om_0$ of a pure sample, and $M_0$
is an effective intensity of direct Coulomb interactions. The latter
approximately equals
$$
M_0 \approx \frac{\pi e^2}{k_F^2}\; \ln \left | 1 + 4
\left ( \frac{k_F}{\kappa}\right )^2 \right | \; ,
$$
where $k_F$ is a Fermi momentum of charge carriers, $\kappa^{-1}$ is a
screening radius, for which $\kappa^2\approx 4m_0e^2(3\rho/\pi)^{1/3}$,
and $m_0$, $e$, and $\rho$ are mass, charge, and density of carriers.

Keeping in mind inequality (42), we set $J_{ij}>0$ and $b_{ij}>0$. Then the
condition for the existence of superconductivity reads
\be
\label{53}
\frac{|\al|^2}{\om_0^2}\; - w M_0 > 0 \; .
\ee
This differs from the Bardeen-Cooper-Schriefer criterion for superconductivity
[63] by the presence of the superconducting phase probability $w$, which
essentially changes the meaning of Eq. (53). It may happen that Coulomb
interactions are so strong, with $M_0>|\al/\om_0|^2$, that superconductivity
in a pure sample is impossible. However, since $w<1$, criterion (53) may be
valid, which implies the occurrence of superconductivity. In this way, we get
the second conclusion:

\vskip 2mm

{\it Phase separation enables the appearance of superconductivity in a
heterophase sample even if it were impossible in pure-phase matter}.

\vskip 2mm

Defining the dimensionless quantity
\be
\label{54}
\mu^* \equiv M_0\; \frac{\om_0^2}{|\al|^2} \; ,
\ee
condition (53) can be presented as
\be
\label{55}
1 - w\mu^* > 0 \; .
\ee
The parameter (54) is of the order [60] of $\mu^*\sim\om_0/\om_p$, where
$\om_p$ is the ion plasma frequency. For good conductors, $\om_0\ll\om_p$,
hence $\mu^*\ll 1$, and inequality (55) is easy to satisfy even for a pure
sample, with $w=1$. For bad conductors, $\om_0\geq \om_p$, so that
$\mu^*\geq 1$. In particular, if $\mu^*>1$, superconductivity cannot arise
in a pure sample, though may appear in phase-separated matter, with $w<1$.
This yields the third conclusion:

\vskip 2mm

{\it Phase separation is crucial for the occurrence of superconductivity in
bad conductors}.

\section{Critical Temperature}

At the critical temperature $T_c$, the gap tends to zero, $\Dlt(\bk)\ra 0$,
hence $E(\bk)\ra\om(\bk)$. Then Eq. (49) becomes
\be
\label{56}
A_{ij}(T_c) = \sum_p \frac{wJ_{ij}}{2\om(\bp)}\; {\rm tanh} \left [
\frac{w\om(\bp)}{2T_c}\right ] \; \chi_i^*(\bp) \chi_j(\bp) \; .
\ee
Using the density of states
\be
\label{57}
N_{ij}(\om) \equiv \sum_p \dlt(\om-\om(\bp)) \chi_i^*(\bp)
\chi_j(\bp) \; ,
\ee
satisfying the normalization
$$
\int_{-\infty}^{+\infty}\; N_{ij}(\om)\; d\om = \dlt_{ij} \; ,
$$
Eq. (56) can be written as
\be
\label{58}
A_{ij}(T_c) = w J_{ij} \int_{-\infty}^{+\infty} \;
\frac{N_{ij}(\om)}{2\om}\; {\rm tanh}\left ( \frac{w\om}{2T_c}
\right ) \; d\om \; .
\ee
The density of states (57), with the standard replacement of summation by
integration
$$
\sum_{p\in{\cal B}} \longrightarrow \frac{1}{\rho} \int_{\cal B} \;
\frac{d\bp}{(2\pi)^3} \; ,
$$
where ${\cal B}$ implies the Brillouin zone, transforms to
\be
\label{59}
N_{ij}(\om) = \frac{1}{\rho} \int_{\cal B} \; \dlt(\om-\om(\bp))\;
\chi_i^*(\bp)\; \chi_j(\bp) \; \frac{d\bp}{(2\pi)^3} \; .
\ee
Assuming, as usual, that the density of states $N_{ij}(\om)$ is the largest
on the Fermi surface and fastly decreases after $\om>\tilde\om_0$, Eq. (58)
can be reduced to
\be
\label{60}
A_{ij}(T_c) = wJ_{ij} N_{ij}(0) \int_0^{\tilde\om_0}\;
\frac{1}{\om}\; {\rm tanh}\left ( \frac{w\om}{2T_c}\right ) \; d\om \; .
\ee
Introducing the coupling matrix $\hat\lbd$, with the elements
\be
\label{61}
\lbd_{ij} \equiv N_{ij}(0) \frac{|\al|^2}{\om_0^2}\; b_{ij} \; ,
\ee
effective coupling matrix $\hat\Lambda$, with
\be
\label{62}
\Lambda_{ij} \equiv w J_{ij} N_{ij}(0) = (1 - w\mu^*)\lambda_{ij} \; ,
\ee
and the characteristic integral
\be
\label{63}
I_c \equiv \int_0^1 \; \frac{1}{x}\; {\rm tanh}\left (
\frac{w^{3/2}\om_0}{2T_c}\; x \right ) \; dx \; ,
\ee
we can present Eq. (60) as
\be
\label{64}
A_{ij}(T_c) = I_c \Lambda_{ij} \; .
\ee
Substituting this into condition (50) gives the equation
\be
\label{65}
det (\hat 1 - I_c \hat\Lbd ) = 0
\ee
for the critical temperature $T_c$.

If $\Lbd_{ij}$ is diagonal, then $T_c$ is defined by the largest
$\Lbd_{ii}$. However, in general, for anisotropic superconductors,
$\Lbd_{ij}$ is not diagonal, and $\Lbd_{ij}\neq 0$ for $i\neq j$. The
latter means that, generally, the gap (47) is presented by a mixture
of waves of different symmetry. There is a very nontrivial relation
between the gap being such a mixture and the magnitude of the critical
temperature, which we describe below.

Let us define an {\it effective coupling} $\Lbd_{eff}$ by the identity
\be
\label{66}
\Lbd_{eff} I_c \equiv 1 - det (\hat 1 - I_c \hat\Lbd) \; .
\ee
Then Eq. (65) for the critical temperature takes the form
\be
\label{67}
\Lbd_{eff} I_c = 1 \; .
\ee
Note that, since, according to definition (63), the integral $I_c>0$
is positive, then $\Lbd_{eff}>0$. From Eq. (67) it follows
\be
\label{68}
\frac{\prt T_c}{\prt\Lbd_{eff}} =
\frac{2T_c^2 I_c}{\om_0 w^{3/2}\Lbd_{eff}I_c'} > 0 \; ,
\ee
where
$$
I_c' \equiv \int_0^1 \; {\rm sech}^2\left (
\frac{w^{3/2}\om_0}{2T_c}\; x\right ) \; dx \; .
$$
Inequality (68) tells us that $T_c$ is higher for larger $\Lbd_{eff}$.
This is valid for all $\Lbd_{eff}>0$ and can be explicitly illustrated
for the particular cases:
$$
T_c \simeq 1.14\; w^{3/2} \om_0\; \exp\left ( -\;
\frac{1}{\Lbd_{eff}}\right ) \qquad (\Lbd_{eff}\ll 1) \; ,
$$
\be
\label{69}
T_c \simeq \frac{1}{2}\; w^{3/2} \om_0\Lbd_{eff} \qquad
(\Lbd_{eff}\gg 1) \; .
\ee

Another important inequality is
\be
\label{70}
\Lbd_{eff} > \max_i \Lbd_{ii} \; ,
\ee
provided that $\Lbd_{ij}\neq 0$ for some $i\neq j$. And, if $\Lbd_{ij}=
\dlt_{ij}\Lbd_{ii}$, then $\Lbd_{eff}=\max_i\Lbd_{ii}$. This property is
easy to explicitly demonstrate for the case, when there are two prevailing
wave symmetries. Thus, if $i=1,2$, then  Eq. (66) yields
\be
\label{71}
\Lbd_{eff} = \frac{1}{2}\left [ \Lbd_{11} +\Lbd_{22} +
\sqrt{(\Lbd_{11}-\Lbd_{22})^2+4\Lbd_{12}^2} \right ] \; .
\ee
Without the loss of generality, we may use the enumeration such that
$\Lbd_{11}>\Lbd_{22}$. As is seen from Eq. (71),
$$
\Lbd_{eff} = \Lbd_{11} \qquad (\Lbd_{12}=0) \; .
$$
In addition,
\be
\label{72}
\frac{\prt\Lbd_{eff}}{\prt\Lbd_{12}^2} =
\frac{1}{\sqrt{(\Lbd_{11}-\Lbd_{22})^2+4\Lbd_{12}^2}} > 0 \; ,
\ee
hence $\Lbd_{eff}$ increases with increasing $|\Lbd_{12}|$. Thus, for
a mixture of waves, when $|\Lbd_{ij}|>0$, where $i\neq j$, the effective
coupling $\Lbd_{eff}$ becomes larger than the maximal $\Lbd_{ii}$. But,
in agreement with Eq. (68), the larger $\Lbd_{eff}$, the higher is the
transition temperature $T_c$. Therefore, we come to the following
conclusion:

\vskip 2mm

{\it Critical temperature for a mixture of gap waves is higher than the
critical temperature related to any pure gap wave from this mixture.}

\vskip 2mm

In superconductors with phase separation, the critical temperature can be
a nonmonotonic function of the superconducting fraction $w$. To show this,
let us consider the case when one of the gap symmetries is prevailing, so
that one of $\Lbd_{ii}$ is essentially larger than other $\Lbd_{ii}$. Let
us denote this maximal $\Lbd_{ii}$ as
\be
\label{73}
\max_i \Lbd_{ii} \equiv (1 - w\mu^*)\lbd \; ,
\ee
where the relation (62) is taken into account. The equation (67) for the
critical temperature can be written as
$$
(1-w\mu^*)\lbd I_c = 1 \; .
$$
The characteristic integral (63) possesses the following asymptotic properties:
$$
I_c \simeq \frac{w^{3/2}\om_0}{2T_c} \qquad (w\ra 0) \; ,
$$
$$
I_c \simeq \ln\left ( 1.14\; \frac{w^{3/2}\om_0}{T_c}\right )
\qquad (T_c\ra 0) \; .
$$
Using these properties, we see that the critical temperature tends to zero
in two limits, if the superconducting fraction tends to zero, when
\be
\label{74}
\frac{T_c}{\om_0} \simeq \frac{1}{2}\; (1-w\mu^*)\lbd w^{3/2}
\qquad (w\ra 0) \; ,
\ee
and also if this fraction tends to a finite value $1/\mu^*$, when
\be
\label{75}
\frac{T_c}{\om_0} \simeq 1.14 w^{3/2}\exp\left\{ -\;
\frac{1}{(1-w\mu^*)\lbd}\right \}
\qquad \left ( w\ra \frac{1}{\mu^*}\right ) \; .
\ee
For good conductors, when $\mu^*\ll 1$, the limit (75) is unachievable. But
for bad conductors, for which $\mu^*\geq 1$, this limit can be achieved.
This can be formulated as another conclusion:

\vskip 2mm

{\it In bad conductors, the critical temperature as a function of the
superconducting fraction $w$ has the bell shape, tending to zero at $w\ra 0$
and at $w\ra 1/\mu^*$.}

\vskip 2mm

To estimate the point where the critical temperature is maximal, we may keep
in mind that experiments with high-temperature superconductors show that only
part of a given sample is in a superconducting phase, this part often being
just a few percent [64,65]. This means that $w\ll 1$ and, hence, the value
$w_{max}$, where $T_c=T_{max}$ is maximal, is also small, $w_{max}\ll 1$.
Taking this into account, from the above equations we obtain
$$
T_{max} \simeq \frac{1}{5}\;\lbd\; \om_0 w^{3/2}_{max} \; , \qquad
w_{max} \simeq \frac{3}{5\mu^*} \; .
$$

Such a bell shape of the critical temperature as a function of doping is
typical of experimental curves for high-temperature cuprate superconductors
[1, 66,67], where the maximal critical temperature occurs at the optimal
doping $0.15$. Assuming that the superconducting fraction is proportional
to the doping, so that $w_{max}\approx 0.15$, we have $\mu^*\approx 4$.
Then the function $T_c(w)$ has a striking similarity with the behaviour
of $T_c$ as a function of doping, studied in experiments with cuprates.
Figure 1 illustrates $T_c(w)$ found by solving numerically Eq. (67).

\section{Density of States}

To concretize the consideration, let us take into account that the
crystalline structure of cuprates is such that the carriers move mainly
in planes, only rarely jumping between the latter, which are separated by
a distance essentially exceeding the mean distance $a$ between lattice sites
on the plane. Neglecting the interplane jumps, one comes to a two-dimensional
motion of carriers on the plane. In the case of such a planar motion, the
single-particle dispersion $\om(\bp)=\om_1(\bp)$, given by Eq. (33), depends
only on two momentum components, say $p_1$ and $p_2$. Then in the following
formulas, it is easy to integrate out the third component $p_3$.

In many cuprates, the Cu and O atoms arrange themselves in a square
lattice with the point group symmetry $C_{4v}$ [25]. This concrete case
will be employed in what follows.

For a square lattice $a\times a$, the Brillouin zone is defined by the
wave vectors $p_\al\in[-\pi/a,\pi/a]$, with $\al=1,2$. It is convenient
to introduce the dimensionless wave vectors $\bk=\{ k_\al\}$, in which
$k_\al\equiv p_\al a$, so that $k_\al\in[-\pi,\pi]$. The point group $C_{4v}$
of a square lattice is characterized by three one-dimensional irreducible
representations labelled as $A_1,\; B_1$, and $B_2$. The representation
$A_1$ is of rank $3$, having three types of symmetries denoted as $s$,
$s^*$, and $s_{xy}$. The representation $B_1$ is of rank $1$, with the
symmetry $d_{x^2-y^2}$. And the irreducible representation $B_2$ is of
rank $1$, with the type of symmetry denoted by $d_{xy}$. The corresponding
basis functions are
$$
\chi_1(\bk) = 1\; ,  \qquad (s)
$$
$$
\chi_2(\bk) = \frac{1}{2}\; ( \cos k_1 + \cos k_2) \; ,
\qquad (s^*)
$$
$$
\chi_3(\bk) = \cos k_1 \cdot \cos k_2 \; , \qquad (s_{xy})
$$
$$
\chi_4(\bk) = \sin k_1 \cdot \sin k_2 \; , \qquad (d_{xy})
$$
$$
\chi_5(\bk) = \frac{1}{2}\; ( \cos k_1 - \cos k_2) \; . \qquad
(d_{x^2-y^2})
$$

For the density of states (57), we have
\be
\label{76}
N_{ij}(\om) = \int_{-\pi}^{\pi} \; \dlt(\om-\om(\bk))\chi_i^*(\bk)
\chi_j(\bk)\; \frac{dk_1\; dk_2}{(2\pi)^2} \; .
\ee
The basis functions $\chi_i^*(\bk)=\chi_i(\bk)$ are real and symmetric with
respect to the inversion of $\bk$, so that $\chi_i(-\bk)=\chi_i(\bk)$. The
single-particle dispersion is also symmetric, $\om(-\bk)=\om(\bk)$. It is
convenient to introduce the dimensionless dispersion
\be
\label{77}
\overline\om(\bk) \equiv \frac{\om(\bk)}{\mu} \; .
\ee
Then the density of states (76) on the Fermi surface becomes
\be
\label{78}
N_{ij}(0) = \frac{1}{\pi^2\mu} \; \int_0^\pi \;
\dlt(\overline\om(\bk)) \chi_i(\bk)\chi_j(\bk)\; dk_1\; dk_2 \; .
\ee

To write down an explicit expression for the single-particle dispersion
$\om(\bk)$, one usually resorts to the tight-binding approximation
[25,39,40,45], in which for a square lattice, one has
\be
\label{79}
\om(\bk) = -t_{eff}(\cos k_1 +\cos k_2) -\mu \; ,
\ee
where $t_{eff}$ is an effective transport parameter. In our case, taking into
account Eq. (33), we see that $t_{eff}$ consists of two parts,
\be
\label{80}
t_{eff} = t_0 + wM_0 \; ,
\ee
the nearest-neighbour hopping integral $t_0$ and the intensity of the repulsive
Coulomb interaction $M_0$. These terms correspond to the transport matrix and
mass operator, respectively. As is well established, both experimentally and
theoretically, a strong Coulomb repulsion is present in all cuprates [25].
With the parameter
\be
\label{81}
t \equiv \frac{t_{eff}}{\mu} = \frac{1}{\mu} \; ( t_0 + w M_0 ) \; ,
\ee
the dimensionless dispersion (77) takes the form
\be
\label{82}
\overline\om(\bk) = - t(\cos k_1 + \cos k_2) -1 \; .
\ee

When the mesoscopic phase separation happens in a superconductor, so that
$w<1$, then, as is seen from Eq. (80), the parameter $t_{eff}$ decreases.
As a result of this, the dispersion (79) becomes softer. Thus, we get an
important conclusion:

\vskip 2mm

{\it Phase separation softens the single-particle dispersion}.

\vskip 2mm

To proceed further, we calculate the density of states (78). For this purpose,
we use the dispersion (82) and note that
$$
\dlt(\overline\om(\bk)) = \frac{\dlt(k_2-k_2(k_1))}{|t\sin k_2(k_1)|} \; ,
$$
where the function $k_2(k)$ is defined by the equation
$$
\cos k_2(k) = -\; \frac{1}{t}\; - \cos k \; .
$$
Then the density of states (78) can be transformed to
\be
\label{83}
N_{ij}(0) = \frac{1}{\pi^2\mu} \;  \int_{k_0}^\pi \;
\frac{\vp_i(k)\vp_j(k)\; dk}{\sqrt{t^2-(1+t\cos k)^2}} \; ,
\ee
where
$$
\vp_i(k) \equiv \chi_i(k,k_2(k)) \; , \qquad k_0\equiv {\rm arc cos}
\left ( 1 -\; \frac{1}{t}\right ) \; ,
$$
$$
0\leq k_0 \leq \pi \qquad \left ( \frac{1}{2}\leq t <\infty \right ) \; .
$$
Explicit expressions for the functions $\vp_i(k)$ are
$$
\vp_1(k) = 1 \; , \qquad \vp_2(k) = -\; \frac{1}{2t}\; , \qquad
\vp_3(k) = -\left ( \frac{1}{t} +\cos k\right )\; \cos k \; ,
$$
$$
\vp_4(k) =\sin k\; \sqrt{1-\left ( \frac{1}{t}+\cos k\right )^2}\; ,
\qquad \vp_5(k) = \frac{1}{2t} + \cos k \; .
$$
Accomplishing in Eq. (83) the change of variables
$$
\psi_i(x) \equiv \vp_i({\rm arccos} x) \qquad (x\equiv \cos k) \; ,
$$
we obtain the form
\be
\label{84}
N_{ij}(0) = \frac{1}{\pi^2\mu t} \;
\int_{-1}^{1-1/t} \; \frac{\psi_i(x)\psi_j(x)}{\psi_4(x)} \; dx \; ,
\ee
in which
$$
\psi_1(x) = 1 \; , \qquad \psi_2(x) = - \; \frac{1}{2t} \; , \qquad
\psi_3(x) = -\left ( \frac{1}{t}+x\right ) \; x \; ,
$$
$$
\psi_4(x) = \sqrt{(1-x^2)\left [ 1 -\left ( \frac{1}{t}+x\right )^2
\right ]} \; , \qquad
\psi_5(x) = \frac{1}{2t}+x \; .
$$
The value of the quantity $N_{ij}(0)$ plays an important role in defining
the coupling parameters (61) and (62). The larger is the density of states
$N_{ii}(0)$, the more profitable is the occurrence of the related gap
symmetry labelled by the index $i$.

Let us analyse the behaviour of Eq. (84) as a function of $t$ changing from
$t=1/2$ to larger $t>1/2$. For convenience, we consider the dimensionless
quantity
\be
\label{85}
D_{ij} \equiv \pi \; \mu N_{ij}(0) \; ,
\ee
normalized by means of
$$
N_{11}(0) = N_{22}(0) = \frac{1}{\pi\mu} \qquad
\left ( t = \frac{1}{2} \right ) \; .
$$
At $t=1/2$, integral (84) can be calculated analytically, giving
$$
D_{11}=D_{13}=D_{22}=D_{33} = 1 \; , \qquad
D_{14}=D_{24}=D_{34}=D_{44}=D_{55} = 0 \; ,
$$
$$
D_{12}=D_{23} = - 1 \; , \qquad \left ( t = \frac{1}{2}\right ) \; .
$$
Also, for all $t\geq 1/2$, one has
$$
D_{15}=D_{25}=D_{35} = D_{45} = 0 \; .
$$
For $t>1/2$, we calculated the integral (84) numerically. The corresponding
densities of states are shown in Fig. 2. Among all $D_{ii}$, only $D_{11}$ and
$D_{22}$ monotonically decrease, all other $D_{ii}$ are nonmonotonic functions
of $t$. Comparing the density of states $D_{ii}$ for different $i=1,2,3,4,5$,
we have the following. Since at $t=1/2$, only $D_{11}=D_{22}=D_{33}=1$ are
nonzero, while $D_{44}=D_{55}=0$, thence only the $s,\; s^*$, and $s_{xy}$
waves can exist. But at $t=1$, we get $D_{11}\approx 0.6$, $D_{22}=D_{44}
\approx 0.2$, $D_{33}\approx 0 $, $D_{55}\approx 0.1$, which tells that now
the gap symmetries $s,\; s^*$, $d_{xy}$, and $d_{x^2-y^2}$ may exist, while
the $s_{xy}$ wave disappears. With increasing $t$, the density $D_{55}$
increases. For instance, at $t=3$, we have $D_{11}\approx 0.3$, $D_{22}$,
and $D_{33}$ are close to zero, $D_{44}=D_{55}\approx 0.1$. Therefore, here
the probable symmetries are $s$, $d_{xy}$, and $d_{x^2-y^2}$. For $t>3$, the
highest densities are $D_{11}$, diminishing below $0.3$, and $D_{55}\approx
0.1$, all other densities being smaller. In this way, for $t>3$ the most
probable symmetries are $s$ and $d_{x^2-y^2}$.

To estimate the magnitude of $t$ for high-temperature superconductors, we may
take the values of parameters typical of cuprates [25,39,40,45], that is,
$t_0\approx (0.5-1)$ eV, $M_0\approx 1$ eV, and $\mu\approx 0.5$ eV. Then the
parameter $t$, defined in Eq. (81), with $w\approx 1$, equals $t\approx 3-4$.
In this region of $t$, the largest densities are $D_{11}$ and $D_{55}$, hence
the symmetries $s$ and $d_{x^2-y^2}$ are preferable. But when phase separation
occurs, the superconducting fraction $w$ becomes less than unity, which
diminishes the value of $t$. The decrease of $t$ suppresses the density of
states $D_{55}$ and enhances $D_{11}$, which means that the relative weights
of $s$ and $d_{x^2-y^2}$ symmetries are changing. This can be formulated as
the following conclusion:

\vskip 2mm

{\it Mesoscopic phase separation suppresses the contribution of $d$-wave
superconductivity and enhances that of $s$-wave superconductivity}.

\vskip 2mm

The value of the superconducting fraction $w$ depends on such parameters as
temperature, pressure, or external magnetic fields. Therefore the relative
contribution of different wave symmetries will be varying under the action
of these parameters. This can explain why in different experiments one
observes alternately the dominance of either $s$ or $d$ gap symmetries.

\section{Summary}

We have studied the main properties of superconductors, such as cuprates,
exhibiting two principal features common for these high-temperature
superconductors, mesoscopic phase separation and anisotropic gap symmetry.
Interplay between these two phenomena is investigated by means of a
model incorporating the presence of mesoscopic phase separation into the
randomly distributed regions of superconducting and normal phases. The
following general conclusions are obtained:

\begin{enumerate}
\item
Mesoscopic phase separation in superconductors can be thermodynamically stable
only in the presence of repulsive Coulomb interactions.

\item
Phase separation enables the appearance of superconductivity in a heterophase
sample even if it were impossible in pure-phase matter.

\item
Phase separation is crucial for the occurrence of superconductivity in bad
conductors.

\item
Critical temperature for a mixture of gap waves is higher than the critical
temperature related to any pure gap wave from this mixture.

\item
In bad conductors, the critical temperature as a function of the
superconducting fraction has the bell shape.

\item
Phase separation softens the single-particle energy dispersion.

\item
Mesoscopic phase separation suppresses the contribution of $d$-wave
superconductivity and enhances that of $s$-wave superconductivity.

\end{enumerate}

These conclusions are in good qualitative agreement with experiments for
high-temperature superconductors. Since in colossal magnetoresistance
materials there also occurs the phenomenon of mesoscopic phase separation
[8--10,68,69], such materials may possess some of the features described
in this paper. Moreover, the mesoscopic phase separation is a rather
general phenomenon appearing in different kinds of condensed matter and
it can be described in the frame of the general theory [5,70], which was
used in this paper for studying the general properties of anisotropic
phase-separated superconductors.

\vskip 5mm

{\bf Acknowledgement}

\vskip 2mm

We are grateful to V. Ivanov for many useful discussions. Financial support
from the German Research Foundation (grant Be 142/72-1) is appreciated.

\newpage

{\large{\bf Appendix. Averaging over Phase Configurations}}

\vskip 5mm

Let the considered sample occupy in the real space $\{\br\}$ a region
$\Bbb{V}$ having the volume $V\equiv\int_{\Bbb{V}} d\br$. Assume that
mesoscopic phase separation occurs in the sample, so that it is filled by
several thermodynamic phases, which are enumerated by an index $\nu$. Say,
$\nu=1$ corresponds to superconducting phase, while $\nu=2$, to normal
phase. At each given instant of time, distinct phases are located at
different spatial regions, which can also be labelled by the phase index
$\nu$. This means that the total sample volume $\Bbb{V}$ can be divided
into subregions $\Bbb{V}_\nu$, filled by the related phases. A family
$\{\Bbb{V}_\nu\}$ of subregions $\Bbb{V}_\nu$ forms an orthogonal
covering of $\Bbb{V}$. This covering can be characterized by a family
$$
\xi \equiv \{ \xi_\nu(\br)|\; \br\in\Bbb{V} \}
$$
of the manifold indicator functions
\begin{eqnarray}
\xi_\nu(\br) \equiv \left \{ \begin{array}{cc}
1, & \br \in \Bbb{V}_\nu \\
\nonumber
0,& \br \not\in \Bbb{V}_\nu \; .
\end{array} \right.
\end{eqnarray}
The family $\xi$ uniquely defines a phase configuration in the
real-space volume $\Bbb{V}$.

From the physical point of view, distinct thermodynamic phases possess
different properties, because of which such phases can be distinguished
from each other. For example, the phases can be distinguished by their
order parameters. In the case of superconductor, a convenient order
parameter can be chosen as the anomalous average
$$
\eta_\nu(\br) \equiv \; <\psi_\nu(\br)\psi_\nu(\br)> \;
\xi_\nu(\br) \; ,
$$
where $\psi_\nu(\br)$ is a field operator of the field of carriers,
when the latter are in the phase $\nu$, and $\xi_\nu(\br)$ is the
related indicator function. For the superconducting regions, one has
$\eta_1(\br)\neq 0$, provided that $\br\in\Bbb{V}_1$, while for the
normal parts, one has $\eta_2(\br)\equiv 0$, with $\br\in\Bbb{V}_2$.
The anomalous average, as is known, is directly related to the gap
in the spectrum of excitations. Thus, different phases could be
distinguished by the existence or absence of the gap in the spectrum.
For a while, when dealing with a nonuniform sample consisting of
separate phases, the anomalous average, depending on the spatial
variable $\br$, is a more convenient order parameter.

Mesoscopic phase separation occurs in the real space in a random
way, which means that the locations and shapes of the phase subregions
$\Bbb{V}_\nu$ are random. Then observable quantities should be
determined with averaging over these random phase configurations.
Since each phase configuration is uniquely defined by the set $\xi$
of the manifold indicator functions, it is necessary to describe an
ensemble $\{\xi\}$ of all possible sets $\xi$, corresponding to all
possible phase configurations. For this purpose, we introduce an
orthogonal subcovering $\{\Bbb{V}_{\nu i}\}$ of each region
$\Bbb{V}_\nu$, such that
$$
\Bbb{V}_\nu = \cup_{i=1}^{n_\nu} \Bbb{V}_{\nu i} \; , \qquad
\Bbb{V}_{\mu i}  \cap \Bbb{V}_{\nu j} =\dlt_{\mu\nu}\; \dlt_{ij}
\Bbb{V}_{\nu i} \; .
$$
To each subregion $\Bbb{V}_{\nu i} $, we ascribe a vector $\ba_{\nu i}
\in\Bbb{V}_{\nu i}$, called the center, playing the role of a local
center of coordinates, so that moving $\ba_{\nu i}$ implies a congruent
motion of $\Bbb{V}_{\nu i}$. The introduced subcovering is uniquely
characterized by a family of the indicator functions
\begin{eqnarray}
\xi_{\nu i}(\br-\ba_{\nu i}) \equiv \left \{ \begin{array}{cc}
1, & \br \in \Bbb{V}_{\nu i} \\
\nonumber
0,& \br \not\in \Bbb{V}_{\nu i} \; ,
\end{array} \right.
\end{eqnarray}
with the property
$$
\sum_{i=1}^{n_\nu} \xi_{\nu i}(\br-\ba_{\nu i}) = \xi_\nu(\br) \; .
$$
By moving the centers $\ba_{\nu i}$ and changing the measure of
$\Bbb{V}_\nu$, it is possible to construct various phase configurations.
To explicitly realize the averaging over these configurations, we define
the differential functional measure
$$
D\xi \equiv \lim_{\{ n_\nu\ra\infty\} } \;
\prod_\nu\; \prod_{i=1}^{n_\nu}
\frac{d\ba_{\nu i}}{V}\; \dlt\left ( \sum_\nu x_\nu - 1 \right )
\; \prod_\nu \; d x_\nu \; ,
$$
in which
$$
x_\nu \equiv \frac{1}{V} \; \int_{\Bbb{V}}\; \xi_\nu(\br)\; d\br \; .
$$
This measure, with varying $\ba_{\nu i}\in\Bbb{V}$ and $x_\nu\in[0,1]$,
induces a topology on the manifold $\{\xi\}$, which results in the
topological configuration space ${\cal X}\equiv\{\xi|\;{\cal D}\xi\}$,
composed of all admissible phase configurations.

For each fixed phase configuration, the observable quantities, represented
by Hermitian operators, depend on the given configuration and have the
structure
$$
A(\xi) =\oplus_\nu\; A_\nu(\xi_\nu) \; ,
$$
being defined on the fiber space ${\cal Y}=\otimes_\nu{\cal H}_\nu$, whose
fibering yields the weighted Hilbert spaces ${\cal H}_\nu$. The dependence
of the operators of observable quantities on the indicator functions,
marking the space filled by the corresponding phases, naturally comes from
the identity
$$
\int_{\Bbb{V}_\nu} \; d\br = \int_\Bbb{V} \; \xi_\nu(\br) \; d\br\; .
$$
This identity is employed when representing the operators through the
integrals of their operator densities:
$$
\hat A(\xi) = \int \hat A(\xi,\br)\; d\br =
\oplus_\nu\; \hat A_\nu(\xi_\nu) $$
$$
\hat A(\xi,\br) = \oplus_\nu \hat A_\nu(\xi_\nu,\br) \; , \qquad
\hat A_\nu(\xi_\nu) = \int \hat A_\nu(\xi_\nu,\br)\; d\br \; .
$$
Here and everywhere in the paper, we denote, for simplicity, the integration
over the whole sample as
$$
\int d\br \equiv \int_\Bbb{V}\; d\br \; .
$$
For example, let us write down the Hamiltonian density
$$
\hat H_\nu(\xi_\nu,\br) =\xi_\nu(\br) \psi_\nu^\dgr(\br) \hat K_\nu(\br)
\psi_\nu(\br) + \frac{1}{2} \;
\int \xi_\nu(\br)\xi_\nu(\br')\psi_\nu^\dgr(\br)\psi_\nu^\dgr(\br')
\hat V_\nu(\br,\br') \psi_\nu(\br')\psi_\nu(\br)\; d\br \; d\br' \; ,
$$
in which $\hat K_\nu(\br)$ is an operator of kinetic energy and
$\hat V_\nu(\br,\br')$ is an effective interaction. Analogously, the
number-of-particle operator density is
$$
\hat N_\nu(\xi_\nu,\br) =
\xi_\nu(\br) \psi_\nu^\dgr(\br)\psi_\nu(\br) \; .
$$
The field operators here are assumed to be columns with respect to spin
indices.

A nonuniform system, composed of several thermodynamic phases, must be
described by the quasiequilibrium (or locally equilibrium) Gibbs ensemble,
with a statistical operator proportional to $e^{-\hat X(\xi)}$, where
$$
\hat X(\xi) \equiv \int \bt(\xi,\br) \left [ \hat H(\xi,\br) -
\mu(\xi,\br)\hat N(\xi,\br) \right ]\; d\br \; .
$$
Here the local inverse temperature $\bt(\xi,\br)$ and the local chemical
potential $\mu(\xi,\br)$ model the system nonuniformity corresponding to
a given phase configuration characterized by a set $\xi$. The natural
thermodynamic potential for a quasiequilibrium system, with random $\xi$,
is
$$
Q= -\ln\; {\rm Tr}_{{\cal Y}} \; \int e^{-\hat X(\xi)}\; D\xi \; .
$$
Assuming, as usual, the existence of the thermodynamic limit, and
accomplishing the averaging over phase configurations, which is characterized
by the differential measure $D\xi$, it is possible to prove [5] that the
thermodynamic potential $Q$ reduces to the form
$$
Q = -\ln\; {\rm Tr}_{{\cal Y}} \; e^{-\bt\tilde H} \; ,
$$
with an effective renormalized Hamiltonian $\tilde H=\oplus_\nu H_\nu$, in
which
$$
H_\nu = \int \left [ \hat H_\nu(w_\nu,\br) -
\mu \hat N_\nu(w_\nu,\br) \right ] \; d\br \; ,
$$
and where the average inverse temperature and average chemical potential,
respectively, are
$$
\bt = \int \bt(\xi,\br) \; D\xi \; , \qquad
\mu = \int \mu(\xi,\br)\; D\xi \; .
$$
Then the potential
$$
Q =\bt \Om
$$
is simply related to the Gibbs grand potential $\Om$. The phase probabilities
$w_\nu$ are defined as the minimizers of either $Q$ or $\Om$.

In this way, after averaging over heterophase configurations, we come
to the description of the system by means of a renormalized Hamiltonian,
containing only averaged quantities and not involving anymore random phase
distributions that have been averaged out. All expressions throughout the
paper correspond to averaged quantities resulting from the described
procedure of heterophase averaging. A complete and detailed mathematical
foundation for this averaging procedure is given in reviews [5,70].

\newpage

\begin{center}
{\large{\bf Figure Captions}}
\end{center}

\vskip 2cm

{\bf Fig. 1}. Critical temperature $T_c$, in units of $\om_0$, as a function
of $w$ for $\mu^*=4$ and different couplings: $\lbd=1$ (dotted line), $\lbd=5$
(dashed line), and $\lbd=10$ (solid line).

\vskip 2cm

{\bf Fig. 2}. Density of states $D_{ii}$ as a function of the effective
transport parameter $t$: $D_{11}$ (upper dashed-double-dotted line), $D_{22}$
(dashed-dotted line), $D_{33}$ (dotted line), $D_{44}$ (dashed line), and
$D_{55}$ (solid line).


\newpage

\begin{references}
\bibitem{1}
J.C. Phillips, Physics of High-T$_c$ Superconductors (Academic, Boston, 1989).

\bibitem{2}
G. Benedek and K.A. M\"uller, editors, {\it Phase Separation in Cuprate
Superconductors} (World Scientific, Singapore, 1992).

\bibitem{3}
E. Sigmund and K.A. M\"uller, editors, {\it Phase Separation in Cuprate
Superconductors} (Springer, Berlin, 1994).

\bibitem{4}
S.A. Kivelson et al., Rev. Mod. Phys. {\bf 75}, 1201 (2003).

\bibitem{5}
V.I. Yukalov, Phys. Rep. {\bf 208}, 395 (1991).

\bibitem{6}
Y.L. Khait, {\it Atomic Diffusion in Solids} (Scitec, Zurich, 1997).

\bibitem{7}
A.J. Coleman, in {\it Phase Transitions and Self-Organization in Electronic
and Molecular Networks}, edited by J.C. Phillips and M.F. Thorpe (Kluwer,
New York, 2001), p. 23.

\bibitem{8}
L.P. Gorkov, Phys. Usp. {\bf 168}, 665 (1998).

\bibitem{9}
M.B. Salamon and M. Jaime, Rev. Mod. Phys. {\bf 73}, 583 (2001).

\bibitem{10}
E. Dagotto, T. Hotta, and A. Moreo, Phys. Rep. {\bf 344}, 1 (2001).

\bibitem{11}
V.I. Yukalov, Ferroelectrics {\bf 82}, 11 (1988).

\bibitem{12}
Y. Yamada {\it et al}., Ferroelectrics {\bf 240}, 363 (2000).

\bibitem{13}
L.P. Gorkov, J. Supercond. {\bf 14}, 365 (2001).

\bibitem{14}
Y.L. Khait, R. Beserman, A. Chack, R. Weil, and W. Beyer, Appl. Phys.
Lett. {\bf 81}, 3347 (2002).

\bibitem{15}
V.I. Yukalov, Phys. Rev. E {\bf 65}, 056118 (2002).

\bibitem{16}
A.S. Shumovsky and V.I. Yukalov, Phys. Dokl. {\bf 27}, 709 (1982).

\bibitem{17}
V.I. Yukalov, {\it On the Model of Heterophase Superconductor} (JINR
Commun. E17-85-114, Dubna, 1985).

\bibitem{18}
J.G. Bednorz and K.A. M\"uller, Z. Phys. B {\bf 64}, 189 (1986).

\bibitem{19}
L.P. Gorkov and A.V. Sokol, J. Exp. Theor. Phys. Lett. {\bf 46}, 420 (1987).

\bibitem{20}
Y.L. Khait, Z. Phys. B {\bf 71}, 7 (1988).

\bibitem{21}
V.J. Emery, S.A. Kivelson, and H.Q. Linn, Phys. Rev. Lett. {\bf 64},
475 (1990).

\bibitem{22}
V.I. Yukalov, Int. J. Mod. Phys. B {\bf 6}, 91 (1992).

\bibitem{23}
A.J. Coleman, E.P. Yukalova, and V.I. Yukalov, Physica C {\bf 243},
76 (1995).

\bibitem{24}
D.J. Van Harlingen, Rev. Mod. Phys. {\bf 67}, 515 (1995).

\bibitem{25}
C.C. Tsuei and J.R. Kirtley, Rev. Mod. Phys. {\bf 72}, 969 (2000).

\bibitem{26}
A. Bhattacharya {\it et al}., Phys. Rev. Lett. {\bf 82}, 3132 (1999).

\bibitem{27}
J. Demsar {\it et al.}, Phys. Rev. Lett. {\bf 82}, 4918 (1999).

\bibitem{28}
Q. Li {\it et al.}, Phys. Rev. Lett. {\bf 83}, 4160 (1999).

\bibitem{29}
V.V. Kabanov, J. Demsar, B. Podobnik, and D. Mihailovic, Phys. Rev. B
{\bf 59}, 1497 (1999).

\bibitem{30}
C.C. Tsuei {\it et al.}, Phys. Rev. Lett. {\bf 73}, 593 (1994).

\bibitem{31}
C. Kendziora, R.J. Kelley, and M. Onellion, Phys. Rev. Lett. {\bf 77},
727 (1996).

\bibitem{32}
M. Covington {\it et al.}, Phys. Rev. Lett. {\bf 79}, 277 (1997).

\bibitem{33}
K.A. Kouznetsov {\it et al.}, Phys. Rev. Lett. {\bf 79}, 3050 (1997).

\bibitem{34}
K. Krishana, N.P. Ong, Q. Li, G.D. Gu, and N. Koshizuka, Science
{\bf 277}, 83 (1997).

\bibitem{35}
H. Strikanth {\it et al.}, Phys. Rev. B {\bf 55}, 14733 (1997).

\bibitem{36}
R.A. Klemm, C.T. Rieck, and K. Scharnberg, Phys. Rev. B {\bf 58},
1051 (1998).

\bibitem{37}
R.A. Klemm {\it et al.}, Phys. Rev. B {\bf 58}, 14203 (1998).

\bibitem{38}
E.G. Maksimov, Phys. Usp. {\bf 170}, 1033 (2000).

\bibitem{39}
G. Kotliar, Phys. Rev. B {\bf 37}, 3664 (1988).

\bibitem{40}
Q.P. Li, B.E.C. Koltenbah, and R. Joynt, Phys. Rev. B {\bf 48},
437 (1993).

\bibitem{41}
C. O'Donovan, D. Branch, J.P. Carbotte, and J.S. Preston, Phys. Rev. B
{\bf 51}, 6588 (1995).

\bibitem{42}
A.A. Abrikosov, Phys. Rev. B {\bf 51}, 11955 (1995).

\bibitem{43}
A.A. Abrikosov, Phys. Rev. B {\bf 52}, 15738 (1995).

\bibitem{44}
C. O'Donovan and J.P. Carbotte,  Phys. Rev. B {\bf 52}, 16208 (1995).

\bibitem{45}
R. Combescot and X. Leyronas, Phys. Rev. Lett. {\bf 75}, 3732 (1995).

\bibitem{46}
M.B. Walker, Phys. Rev. B {\bf 53}, 5835 (1996).

\bibitem{47}
A.A. Abrikosov, Phys. Rev. B {\bf 53}, 8910 (1996).

\bibitem{48}
M.B. Walker and J. Luettmer-Strathmann, Phys. Rev. B {\bf 54}, 588 (1996).

\bibitem{49}
W. Xu, W. Kim, Y. Ren, and C.S. Ting, Phys. Rev. B {\bf 54}, 12693 (1996).

\bibitem{50}
M. Fogelstr\"om, D. Rainer, and J.A. Sauls, Phys. Rev. Lett. {\bf 79},
281 (1997).

\bibitem{51}
G. Presti and M. Palumbo, Phys. Rev. B {\bf 55}, 8430 (1997).

\bibitem{52}
I. Sch\"urrer, E. Schachinger, and J.P. Carbotte, Physica C {\bf 303}, 287
(1998).

\bibitem{53}
A.J. Coleman and V.I. Yukalov, {\it Reduced Density Matrices} (Springer, Berlin,
2000).

\bibitem{54}
V.I. Yukalov, Physica A {\bf 310}, 413 (2002).

\bibitem{55}
H. Fr\"olich, Phys. Rev. {\bf 79}, 845 (1950).

\bibitem{56}
N.N. Bogolubov, J. Exp. Theor. Phys. {\bf 34}, 58 (1958).

\bibitem{57}
G.M. Eliashberg, J. Exp. Theor. Phys. {\bf 38}, 966 (1960).

\bibitem{58}
G.M. Eliashberg, J. Exp. Theor. Phys. {\bf 39}, 1437 (1960).

\bibitem{59}
V. Ivanov, Philos. Mag. B {\bf 76}, 697 (1997).

\bibitem{60}
D. Pines, {\it Elementary Excitations in Solids} (Benjamin, New York, 1963).

\bibitem{61}
V.I. Yukalov, Chem. Phys. Lett. {\bf 229}, 239 (1994).

\bibitem{62}
V.I. Yukalov, Physica A {\bf 213}, 500 (1995).

\bibitem{63}
J. Bardeen, L.N. Cooper, and J.R. Schrieffer, Phys. Rev. {\bf 108}, 1175 (1957).

\bibitem{64}
J. Tholence {\it et al.}, Phys. Lett. A {\bf 184}, 215 (1994).

\bibitem{65}
A. Lappas {\it et al}., Physica B {\bf 194}, 353 (1994).

\bibitem{66}
N. Tsuda, K. Nasu, A. Yanase, and K. Siratori, {\it Electronic Conduction in
Oxides} (Springer, Berlin, 1991).

\bibitem{67}
T. Schneider and H. Keller, Phys. Rev. Lett. {\bf 69}, 3374 (1992).

\bibitem{68}
J.P. Renard and M. Velazquez, Eur. Phys. J. B {\bf 34}, 41 (2003).

\bibitem{69}
K. Shimizu, M. Velazquez, J.P. Renard, and A. Revcolevschi, J. Phys. Soc.
Jap. {\bf 72}, 793 (2003).

\bibitem{70}
V.I. Yukalov, Int. J. Mod. Phys. B {\bf 17}, 2333 (2003).

\end{references}
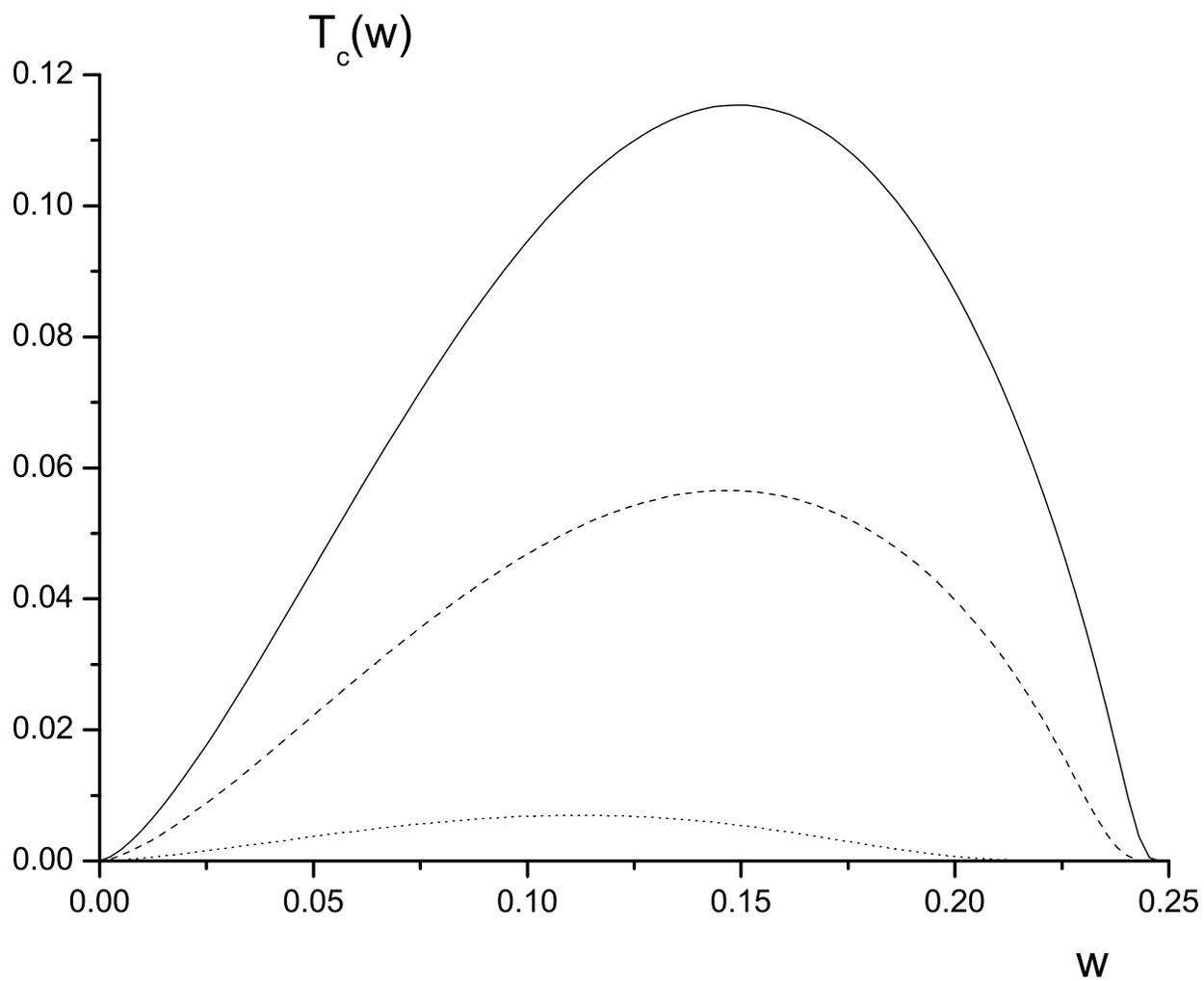
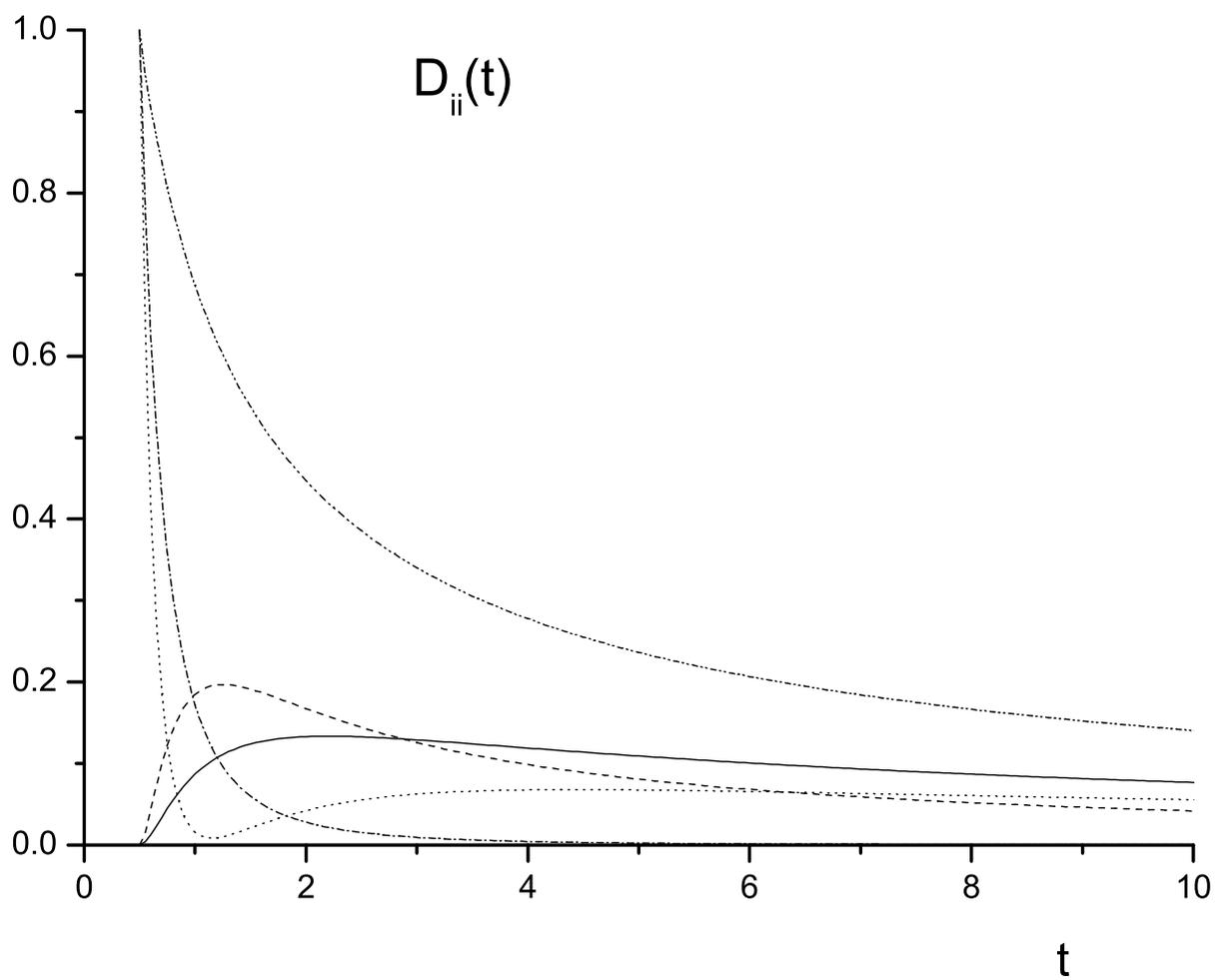
\end{document}